\documentclass[pdflatex,sn-mathphys-num]{sn-jnl}

\usepackage{graphicx}%
\usepackage{multirow}%
\usepackage{amsmath,amssymb,amsfonts}%
\usepackage{amsthm}%
\usepackage{mathrsfs}%
\usepackage[title]{appendix}%
\usepackage{xcolor}%
\usepackage{textcomp}%
\usepackage{manyfoot}%
\usepackage{booktabs}%
\usepackage{algorithm}%
\usepackage{algorithmicx}%
\usepackage{algpseudocode}%
\usepackage{listings}%

\usepackage{graphicx}
\usepackage{dcolumn}
\usepackage{bm}
\usepackage{caption}

\usepackage{graphicx}
\usepackage{ragged2e}
\usepackage{dcolumn}
\usepackage{bm}

\usepackage[most]{tcolorbox}
\usepackage{bbm}
\usepackage{multirow}
\usepackage{amsfonts}
\usepackage{amsthm}
\usepackage{mathrsfs}
\usepackage[title]{appendix}
\usepackage{xcolor}
\usepackage{textcomp}
\usepackage{booktabs}
\usepackage{algorithm}
\usepackage{algorithmicx}
\usepackage{algpseudocode}
\usepackage{listings}

\theoremstyle{plain}

\theoremstyle{definition}

\theoremstyle{remark}

\begin{document}

\title[Signature structure of quadratic response under Zeno–Schur coarse graining in open quantum systems]{Signature structure of quadratic response under Zeno–Schur coarse graining in open quantum systems}

\author*[1]{\fnm{Ansgar} \sur{Pernice}}\email{pernice@gmx.net}



\affil*[1]{\orgaddress{ \city{Kochel am See}, \postcode{82431}, \state{Bavaria}, \country{Germany}}}



\abstract{Quadratic response tensors arise naturally in quantum kinetic descriptions,
such as the quantum linear Boltzmann equation (QLBE), where they encode the
coupled structure of drift and fluctuations beyond simple positive-definite
forms. Motivated by this class of systems, we investigate how such response
structures are modified under monitoring-induced coarse graining.

Within the Gorini--Kossakowski--Sudarshan--Lindblad (GKSL) framework and under
time-scale separation, Zeno elimination of fast degrees of freedom generates
a subtractive renormalization with Schur-complement structure. As a result,
positive definiteness of the response tensor is not preserved: coupling
between slow and rapidly damped sectors can induce negative directions even
when the microscopic tensor is strictly positive.

We formulate a minimal effective flow capturing this mechanism and show that
the competition between Schur-induced compression and anisotropic
perturbations organizes the dynamics into distinct signature sectors. The
resulting structure appears to be robust within the class of models considered and, in appropriate regimes, may be experimentally accessible.

Our results establish a general framework for how quadratic response
structures, as encountered in QLBE-type dynamics, are dynamically reorganized
under Zeno-induced coarse graining.}


\maketitle

\section{Introduction}\label{sec:1}

Open quantum systems subject to continuous monitoring exhibit a rich interplay
between coherent dynamics, dissipation, and measurement backaction.
Their dynamics is commonly described within the framework of completely positive
Markovian semigroups, as formalized by the
Gorini--Kossakowski--Sudarshan--Lindblad (GKSL) equation
\cite{Lindblad1976,Gorini1976}, and extensively developed in the theory of open
quantum systems \cite{BreuerPetruccione}.
In regimes of strong monitoring, the dynamics can undergo a quantum Zeno reduction,
in which rapidly relaxing degrees of freedom are effectively eliminated, yielding a
constrained evolution on a slow manifold.
Such Zeno-reduced dynamics arise naturally under standard assumptions of time-scale
separation and adiabatic elimination, and provide a controlled setting for studying
coarse-grained behavior in monitored systems \cite{Kato1950,Azouit2016}.

More broadly, measurement-induced dynamics has recently attracted considerable
attention in the context of monitored many-body systems, where measurement
backaction can fundamentally alter dynamical behavior and lead to emergent structures
in effective descriptions \cite{LiChenFisher2018,SkinnerRuhmanNahum2019}.
At the same time, driven open quantum systems have been shown to exhibit universal
features at long wavelengths, highlighting the importance of coarse-grained
descriptions beyond microscopic details \cite{Sieberer2025}.

A central question in this context concerns the structure of effective observables
under monitoring-induced coarse graining.
In particular, quadratic response tensors, which characterize fluctuations and
linear response of coarse-grained degrees of freedom, play a central role in
describing long-wavelength dynamics and transport processes.
Such quadratic structures are closely related to information-geometric descriptions
of statistical systems, where local curvature encodes fluctuation response
\cite{Amari2016}.
In standard treatments of fluctuations and transport in statistical and kinetic
theories, positive-definite quadratic forms are typically assumed to remain stable
under coarse-graining~\cite{LandauLifshitzStat,Zwanzig2001}.
However, it is not a priori guaranteed that this property is preserved under
elimination procedures involving measurement backaction and dissipative coupling.

Recent developments in quantum kinetic descriptions, in particular within the
framework of the quantum linear Boltzmann equation (QLBE), indicate that
effective quadratic response tensors of this type arise naturally as
coarse-grained descriptors of monitored dynamics \cite{Vacchini2009,Pernice2026}.
In such settings, the response tensor encodes the structure of drift and
fluctuation relations beyond simple positive-definite covariance forms. This observation motivates treating quadratic response tensors as primary objects of interest, rather than derived quantities, in the analysis of coarse-grained monitored dynamics.

In this work, we identify a structural mechanism by which Zeno reduction
modifies quadratic observables through a subtractive renormalization of
Schur-complement form (Sec.~\ref{sec:2}). Within a class of GKSL dynamics
characterized by time-scale separation, linear slow--fast coupling, and
closure at quadratic order, elimination of fast degrees of freedom induces
a positive semidefinite correction that enters subtractively. As a result,
the effective quadratic response is not constrained to remain positive
definite and can develop negative directions under sufficiently strong
coupling.

To connect this structural mechanism to physically realizable settings, we
construct a minimal realization based on coherence-sensitive monitoring of
coupled modes (Sec.~\ref{sec:3}), which explicitly generates the
off-diagonal structure responsible for the Schur-induced modification.

Building on this result, we analyze the induced dynamics on the space of
quadratic response tensors using a minimal effective coarse-grained model
(Sec.~\ref{sec:4}). Within this framework, the interplay between
elimination-induced isotropic contraction and anisotropic perturbations
organizes the system into distinct signature sectors, classified by the
number of negative eigenvalues. The resulting dynamics exhibits a
pronounced but continuous crossover between regimes, controlled by a
dimensionless ratio comparing anisotropic forcing to Schur-induced damping.

This structure is characterized statistically by analyzing ensembles of
trajectories generated by the effective dynamics. We extract observables
such as the distribution of signature sectors and first-passage times,
revealing a well-defined boundary between regimes and signatures of slowing
down near the crossover. These features indicate that the selection of
signature sectors arises as a robust property of the effective dynamics
within the class of models considered.

Taken together, these results show that the signature of quadratic response
tensors in monitored open quantum systems is not fixed a priori, but can be
dynamically modified by Zeno-induced renormalization. Since such tensors
govern both dynamical response and fluctuation structure, their signature
plays a central role in determining the qualitative behavior of the
effective theory. The possibility of dynamically selected signature
structures thus points to a mechanism by which nontrivial effective
kinematics may emerge in monitoring-induced dynamics.

The central result is therefore that Zeno-induced elimination provides,
within the class of systems considered, a generic mechanism by which
quadratic response tensors lose positive definiteness through a
subtractive Schur-complement renormalization, enabling the emergence of
nontrivial signature structure under coarse graining.

\section{Zeno reduction and Schur renormalization}\label{sec:2}

In the following, ``generic'' refers to structural robustness under variations of coupling matrices and anisotropic perturbations within the
class of Zeno-reduced GKSL dynamics satisfying (i) time-scale separation,
(ii) quadratic closure, and (iii) the existence of a compatible quadratic response
representation, rather than to arbitrary open quantum systems.

We consider monitored open quantum systems described by a
Gorini--Kossakowski--Sudarshan--Lindblad (GKSL) master equation
\begin{equation}
\frac{d\rho}{dt} = \mathcal{L}[\rho],
\end{equation}
where $\mathcal{L}$ generates completely positive Markovian dynamics
\cite{Lindblad1976,Gorini1976}. We focus on regimes of strong continuous
monitoring, in which measurement backaction induces a separation of time
scales between rapidly relaxing and slowly evolving degrees of freedom,
leading to Zeno-type constrained dynamics
\cite{FacchiPascazio2008,Gough2014,WisemanMilburn2010}.

\paragraph{Moment dynamics and slow--fast decomposition.}
To connect the GKSL evolution to effective response structures, we consider
a set of observables $\{O_i\}$ whose expectation values define a vector of
coarse-grained variables
\begin{equation}
x_i = \langle O_i \rangle.
\end{equation}
Under the adjoint GKSL evolution, these variables satisfy
\begin{equation}
\frac{d}{dt} x_i = \langle \mathcal{L}^\dagger O_i \rangle.
\end{equation}
We assume that, within the sector of interest, the dynamics closes linearly,
so that
\begin{equation}
\label{eq:momentDynamics}
\frac{d}{dt}
\begin{pmatrix}
x_s \\
x_f
\end{pmatrix}
=
\begin{pmatrix}
K_{ss} & K_{sf} \\
K_{fs} & K_{ff}
\end{pmatrix}
\begin{pmatrix}
x_s \\
x_f
\end{pmatrix},
\end{equation}
where $x_s$ and $x_f$ denote slow and fast components, respectively.
We assume that the fast sector is dynamically stable,
$\mathrm{Re}\,\mathrm{spec}(K_{ff}) < 0$.

This structure is closely related to projection-based formulations of
open-system dynamics and irreversibility
\cite{Nakajima1958,Zwanzig1960,Feshbach1958}, as well as to modern treatments
of adiabatic elimination in GKSL systems \cite{Azouit2016,Reiter2012}.

\paragraph{Zeno elimination at the level of moments.}
In the Zeno regime, the fast variables relax rapidly to a quasi-stationary
state determined by the slow sector. Setting $\dot{x}_f \simeq 0$ in
Eq.~\eqref{eq:momentDynamics} yields
\begin{equation}
x_f \simeq -K_{ff}^{-1}K_{fs}x_s.
\end{equation}
Substituting into the slow equation gives the effective reduced dynamics
\begin{equation}
\label{eq:Keff}
\dot{x}_s =
\left(
K_{ss} - K_{sf}K_{ff}^{-1}K_{fs}
\right) x_s.
\end{equation}
This is the standard Schur-complement structure associated with adiabatic
elimination. At the level of density operators, the same structure appears in the
Liouville-space decomposition
\begin{equation}
\frac{d}{dt}
\begin{pmatrix}
\rho_s \\
\rho_f
\end{pmatrix}
=
\begin{pmatrix}
\mathcal{L}_{ss} & \mathcal{L}_{sf} \\
\mathcal{L}_{fs} & \mathcal{L}_{ff}
\end{pmatrix}
\begin{pmatrix}
\rho_s \\
\rho_f
\end{pmatrix},
\end{equation}
leading to the effective generator
\begin{equation}
\label{eq:LeffSchur}
\mathcal{L}_{\mathrm{eff}} =
\mathcal{L}_{ss}
-
\mathcal{L}_{sf}\mathcal{L}_{ff}^{-1}\mathcal{L}_{fs}.
\end{equation}

While Eqs.~\eqref{eq:Keff} and \eqref{eq:LeffSchur} are formally standard,
their consequences for the structure of effective observables are not
fully captured by the reduced generator alone.
In particular, the elimination-induced coupling between slow and fast sectors
introduces a qualitatively distinct modification when observables admit a
quadratic response representation, as we now show.

\paragraph{Emergence of a quadratic response structure.}
To relate the reduced dynamics to quadratic observables, we assume that the
slow-sector dynamics admits a local gradient (or generalized Onsager) form,
\begin{equation}
\label{eq:gradientStructure}
\dot{x} = -\mu \nabla_x \Phi(x),
\end{equation}
where $\mu$ is a positive (mobility) operator and $\Phi$ is a quadratic
response functional
\begin{equation}
\label{eq:responseFunctional}
\Phi(x) = \frac{1}{2} x^{\top} Q x.
\end{equation}
This assumption is satisfied, for example, in linear stochastic systems with
Gaussian fluctuations or in regimes admitting a local Einstein relation (see
Sec.~\ref{sec:6}). We do not assume that all GKSL dynamics admit such a representation; rather, we restrict to regimes where an effective quadratic response description is valid. Under Eq.~\eqref{eq:gradientStructure}, the drift matrix
is given by
\begin{equation}
K = -\mu Q.
\end{equation}
We assume that the mobility operator $\mu$ is positive definite and
compatible with the slow--fast decomposition, in the sense that it does
not introduce leading-order mixing between sectors (e.g.\ it is
block-diagonal to leading approximation). Under this condition, the Schur
reduction of $K$ in Eq.~\eqref{eq:Keff} induces a corresponding transformation
at the level of the quadratic response tensor.

Such quadratic response representations arise naturally once quantum kinetic
generators are restricted to their Brownian or Fokker--Planck sector. In
QLBE-type descriptions, the Markovian generator contains coupled dissipative
drift and fluctuation terms, and its Brownian limit reduces to a linear
Fokker--Planck or Ornstein--Uhlenbeck form with drift matrix $K$ and diffusion
matrix $D$ \cite{Vacchini2009,Risken1989}. If the reduced stationary state is
locally Gaussian,
\begin{equation}
P_{\mathrm{st}}(x) \propto
\exp\!\left[-\frac{1}{2}x^{\top}Qx\right],
\end{equation}
then the corresponding quadratic response tensor is the logarithmic curvature
of the stationary distribution. Under a local fluctuation--dissipation or
Einstein relation, the dissipative drift can be written in Onsager form~\eqref{eq:gradientStructure} so that $K=-\mu Q$ with positive mobility $\mu$. This is precisely the
linearized form of the irreversible sector in the GENERIC/Onsager framework,
where positive kinetic matrices multiply gradients of thermodynamic potentials
\cite{Grmela1997,Ottinger1997}. Thus, in the quantum kinetic regimes relevant
here, quadratic response tensors are not merely formal objects: they organize
the coupled drift--fluctuation structure of the reduced open-system dynamics. The present construction should be understood
as an abstract formulation of this class.

\paragraph{Schur reduction of the quadratic response.}
In a basis in which the mobility is absorbed into a symmetrized form,
the quadratic response tensor inherits the same block structure as the
dynamical generator. Writing
\begin{equation}
\label{eq:quadraticBlock}
Q =
\begin{pmatrix}
A & B \\
B^{\top} & C
\end{pmatrix},
\end{equation}
with $C>0$ corresponding to rapidly damped directions, elimination of the
fast sector yields
\begin{equation}
\label{eq:Qeff}
Q_{\mathrm{eff}} = A - B C^{-1} B^{\top}.
\end{equation}

Although Eq.~\eqref{eq:Qeff} has the algebraic form of a Schur complement,
its physical role here is more specific: it represents the induced
transformation of a quadratic response functional under dynamical
coarse-graining generated by Zeno elimination.
In this sense, the Schur structure is not merely a formal identity, but
encodes how coupling to eliminated degrees of freedom feeds back into the
effective response of the retained variables. While the mathematical structure itself is standard in elimination
procedures, its interpretation as a transformation acting on quadratic
response tensors—and the resulting loss of positive definiteness—has not been
identified as a generic structural consequence of Zeno-reduced open-system
dynamics.

Since $C^{-1} > 0$, the correction term $B C^{-1} B^{\top}$ is positive
semidefinite and enters subtractively. The key structural feature is
therefore that elimination generates a \emph{systematic reduction} of the
quadratic response along directions coupled to fast modes.

\paragraph{Remark on quadratic closure.}
The assumption of closure at quadratic order means that the evolution of the
chosen set of observables is fully determined by linear combinations of their
first moments, without generating higher-order cumulants. This condition is
satisfied, for example, in Gaussian-preserving dynamics, linear bosonic systems,
and certain weakly interacting regimes where higher-order correlations remain
slaved to second moments. In the present work, it should be understood as a
restriction to a class of systems admitting an effective quadratic description,
rather than as a generic property of GKSL dynamics.

\paragraph{Loss of positive definiteness.}
As a consequence of the subtractive structure in Eq.~\eqref{eq:Qeff}, the
effective quadratic response satisfies
\begin{equation}
Q_{\mathrm{eff}} \leq A,
\end{equation}
and positive definiteness is not generally preserved under elimination.
In particular, sufficiently strong coupling between slow and fast sectors
can induce directions $v$ for which
\begin{equation}
v^{\top} Q_{\mathrm{eff}} v < 0,
\end{equation}
even if the microscopic tensor $Q$ is strictly positive definite.

This loss of positivity does not arise from a breakdown of the underlying
GKSL dynamics, but from the structured backaction of eliminated degrees of
freedom on the effective response sector. In particular, it is controlled by
the coupling block $B$, which encodes coherence-sensitive mixing between slow
and fast variables.

The origin of this effect can be traced to coherence-sensitive couplings
between slow and fast degrees of freedom, which have been shown to induce
Zeno-constrained quadratic structures in monitored open systems
\cite{Pernice2026}. In the absence of such couplings ($B = 0$), the correction
vanishes and positivity is preserved. For finite coupling, the subtractive
Schur term therefore provides a systematic and model-independent mechanism
by which elimination modifies the structure of quadratic observables.

Within the class of dynamics considered here—characterized by time-scale
separation, stability of the fast sector, and closure at quadratic order,
together with the existence of a compatible quadratic response
representation—the above mechanism is structural and does not depend on
microscopic details beyond the induced coupling matrices. It thus identifies
a general pathway by which monitoring-induced coarse graining can generate
indefinite quadratic response tensors, and provides the foundation for the
dynamical analysis of signature structure developed in Sec.~\ref{sec:4}.

\section{Minimal realization of Schur-induced instability}
\label{sec:3}

The structural mechanism identified in Sec.~\ref{sec:2} requires a
nonvanishing coupling block $B$ between slow and rapidly eliminated degrees
of freedom. In abstract form this coupling is simply an off-diagonal block of
the quadratic tensor in Eq.~\eqref{eq:quadraticBlock}. Physically, however, it
has a more specific meaning: it measures the sensitivity of the monitored
quadratic observable to coherences between sectors that relax on different
time scales. In this section we show that such a block arises naturally in
coherence-sensitive monitoring of coupled modes.

We consider a set of modes subject to dissipative mixing and continuous
measurement of a quadratic observable. Such monitored dynamics can be
implemented in a broad class of open quantum systems, and closely related
measurement and feedback settings are described in
Refs.~\cite{WisemanMilburn2010,Albarelli2023}. In a suitable basis, the
monitored observable contains both diagonal contributions, selecting a
preferred set of modes, and off-diagonal terms encoding sensitivity to
coherences between them. The latter generate a coupling between slow and
rapidly damped sectors, corresponding to a nonvanishing off-diagonal block
$B$ in Eq.~\eqref{eq:quadraticBlock}.

A natural physical platform for such coherence-sensitive coupled-mode
structures is provided by synthetic photonic dimensions, where discrete
frequency modes can be dynamically coupled and monitored in a controllable
way \cite{Yuan2021,Lin2018,Hu2020,Dutt2022,Dinh2024,Wang2025}. In these
systems, off-diagonal couplings between modes are experimentally tunable and
can therefore serve as a direct realization of the slow--fast mixing encoded
by the block $B$. In the following, however, this platform should be
understood as an illustrative realization rather than a necessary ingredient
of the mechanism.

In the strong-monitoring (Zeno) regime, the fast sector associated with the
block $C$ is rapidly stabilized, and the effective quadratic response is
given by Eq.~\eqref{eq:Qeff}. As discussed in Sec.~\ref{sec:2}, the
Schur-complement contribution is positive semidefinite and enters
subtractively. To make its effect explicit, we consider the quadratic form
evaluated along a direction $v$ in the slow sector,
\begin{equation}
v^{\top} Q_{\mathrm{eff}} v
=
v^{\top} A v - v^{\top} B C^{-1} B^{\top} v,
\label{eq:veff}
\end{equation}
which shows directly how coupling to eliminated degrees of freedom reduces
the effective response along directions overlapping with the image of $B$.

To illustrate the onset of instability, we parametrize the strength of
coherence-sensitive contributions by a dimensionless parameter $\chi$, which
controls the magnitude of the off-diagonal block $B$. For $\chi = 0$, the
observable is purely diagonal, $B = 0$, and Eq.~\eqref{eq:Qeff} reduces to
$Q_{\mathrm{eff}} = A$, preserving positive definiteness. For finite $\chi$,
the induced coupling generates a nonzero correction. Equation~\eqref{eq:veff}
then implies that sufficiently strong coherence sensitivity can produce
directions $v$ for which
\begin{equation}
v^{\top} Q_{\mathrm{eff}} v < 0,
\label{eq:negativityCondition}
\end{equation}
even when the unreduced quadratic tensor is positive definite on the full
slow--fast space.

\begin{figure}[t] \centering \includegraphics[width=\textwidth]{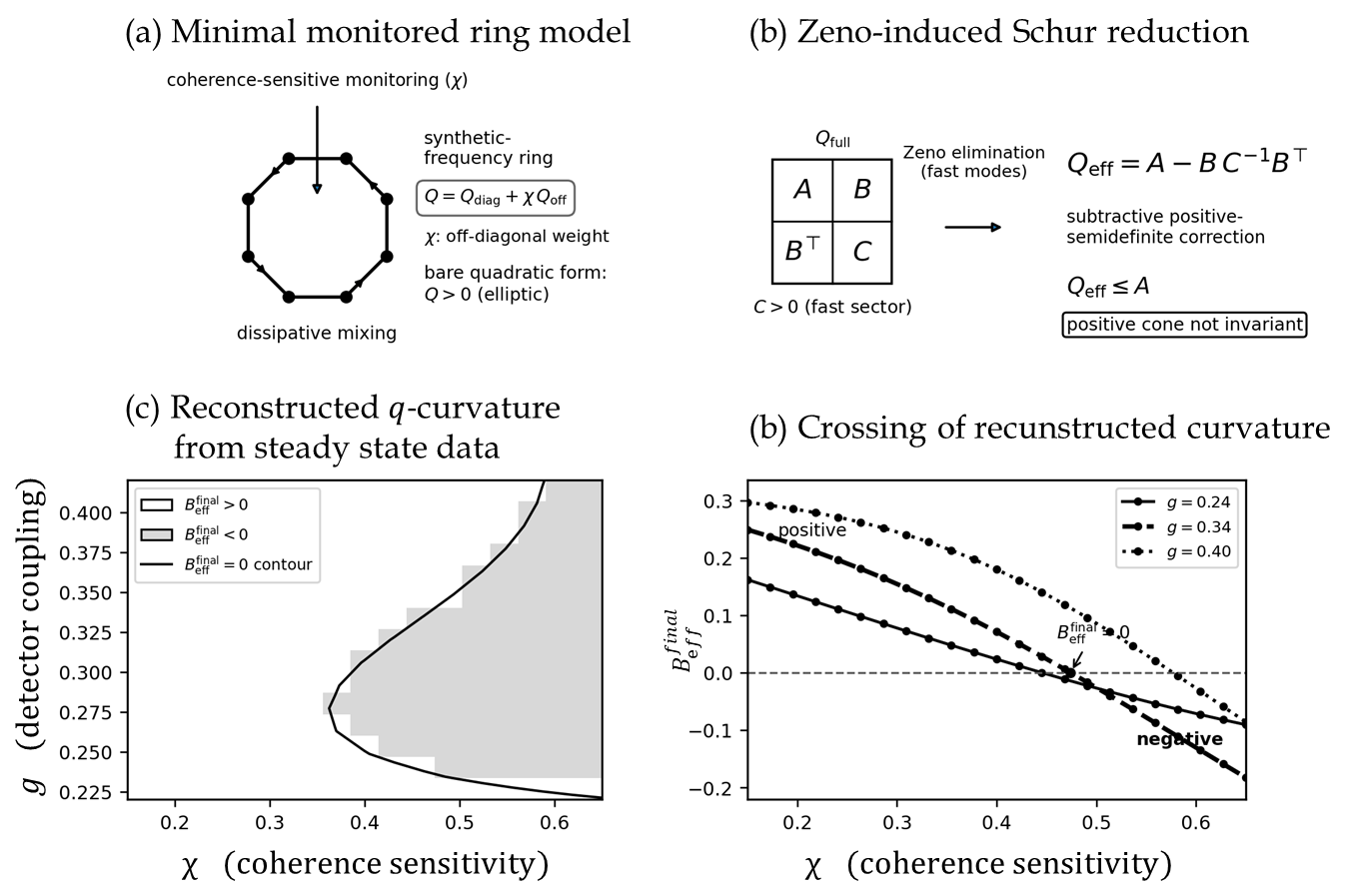} \caption{Zeno-induced Schur renormalization destabilizes positive quadratic structure.
(a) Synthetic-frequency ring with elliptic microscopic dynamics ($Q>0$), subject to coherence-sensitive monitoring and dissipative mixing.
(b) In the strong-measurement regime, Zeno reduction eliminates fast degrees of freedom and induces a Schur-complement renormalization, $Q_{\mathrm{eff}} = A - B C^{-1} B^\top$, yielding a subtractive positive-semidefinite correction and implying that the positive cone is not invariant under coarse graining.
(c) Phase diagram of the reconstructed curvature $B_{\mathrm{eff}}^{\mathrm{final}}$ as a function of coherence sensitivity $\chi$ and detector coupling $g$. A finite connected parameter region with $B_{\mathrm{eff}}^{\mathrm{final}}<0$ emerges, separated by the zero-curvature contour (solid line), demonstrating the breakdown of positive definiteness.
(d) Crossing of $B_{\mathrm{eff}}^{\mathrm{final}}(\chi)$ for different values of $g$, showing a robust sign change and providing operational evidence for the emergence of a negative direction in $Q_{\mathrm{eff}}$.}
\label{fig:schur}
\end{figure}

This mechanism is illustrated in Fig.~\ref{fig:schur}. Panel (a) shows a
representative coupled-mode realization with an initially positive-definite
quadratic structure. Panel (b) displays the corresponding Zeno-induced
elimination of fast degrees of freedom. The resulting Schur-complement
renormalization in Eq.~\eqref{eq:Qeff} subtractively reduces the effective
quadratic response in the retained sector and therefore prevents the positive
cone from being invariant under coarse graining.

Panels (c) and (d) provide an operational characterization of this effect.
In panel (c), the reconstructed effective curvature
$B_{\mathrm{eff}}^{\mathrm{final}}$ is shown as a function of the coherence
sensitivity $\chi$ and detector coupling $g$. A finite connected region with
$B_{\mathrm{eff}}^{\mathrm{final}}<0$ emerges, demonstrating that increasing
coherence-sensitive coupling can drive the reduced system into a regime with
an effective negative direction. Panel (d) shows representative cuts
$B_{\mathrm{eff}}^{\mathrm{final}}(\chi)$ for different values of $g$, where
the crossing of zero gives a direct signature of the instability condition in
Eq.~\eqref{eq:negativityCondition}.

This construction demonstrates explicitly how the abstract block $B$ entering
the Schur complement is generated by a physically controlled feature of the
monitored observable. The emergence of negative directions in
$Q_{\mathrm{eff}}$ is therefore not an arbitrary algebraic possibility, but a
dynamical consequence of coherence-sensitive coupling between retained and
eliminated sectors. The example should be viewed as a minimal realization of
the general Zeno--Schur mechanism, not as a unique microscopic
implementation. The
same structural ingredients appear, however, in more microscopic descriptions such
as the quantum linear Boltzmann equation, where coherence-sensitive couplings
generate effective quadratic response tensors with nontrivial structure~\cite{Pernice2026}.

The role of this section is thus twofold. First, it shows that the
instability mechanism of Sec.~\ref{sec:2} can be realized by experimentally
controllable couplings. Second, it motivates the effective description used
below, in which repeated coarse-graining steps are represented by a flow on
the space of quadratic tensors. In the following section, we analyze how the
competition between Schur-induced renormalization and additional anisotropic
perturbations organizes the system into distinct signature sectors.

\section{Dynamics and signature structure}
\label{sec:4}

Building on the structural mechanism established in Sec.~\ref{sec:2} and its
minimal realization in Sec.~\ref{sec:3}, we now analyze how repeated
Zeno--Schur coarse graining can organize quadratic response tensors into
distinct signature sectors.

As shown in Sec.~\ref{sec:2}, Zeno-induced elimination leads, at the level of
slow observables, to a reduced linear dynamics with Schur-complement structure.
Under the additional assumption of a compatible quadratic response
representation, this induces a corresponding subtractive renormalization of
the quadratic tensor $Q$.

In the following, we do not derive a microscopic renormalization group flow
from the GKSL dynamics. Instead, we construct a \emph{minimal effective
coarse-grained flow} on the space of quadratic tensors. Its purpose is to
capture the structural consequences that are fixed by Zeno--Schur elimination,
while parametrizing residual anisotropic effects not determined by the Schur
term itself. Thus, the results of this section should be read as the dynamics
of an effective model constrained by the mechanism of Sec.~\ref{sec:2}, not as
a complete microscopic derivation.

\paragraph{Effective coarse-grained flow}

The starting point is the Zeno-reduced generator obtained from the slow--fast
decomposition in Sec.~\ref{sec:2}. For a generator with a stable fast sector,
adiabatic elimination to leading nontrivial order is given by
Eq.~\eqref{eq:LeffSchur}, which induces, under the assumptions stated in
Sec.~\ref{sec:2}, a subtractive Schur-type correction to the quadratic
response tensor.

In monitored settings with persistent time-scale separation, such elimination
steps occur repeatedly under coarse-grained observation. Rather than
postulating an ad hoc effective model, we therefore construct a minimal
coarse-grained flow that captures the structural consequences of iterated
Zeno-induced elimination in GKSL dynamics under quadratic closure.

Assuming closure of the reduced dynamics at quadratic order, we parametrize
the coarse-grained state by a symmetric tensor $Q$. The effect of one
coarse-graining step is then represented as an update
\begin{equation}
Q \mapsto Q + \delta Q_{\mathrm{eff}},
\label{eq:QStepGeneral}
\end{equation}
where $\delta Q_{\mathrm{eff}}$ contains the leading contributions consistent
with Zeno-induced elimination.

Guided by the Schur-complement structure derived in Sec.~\ref{sec:2}, the
leading contribution is fixed by the elimination procedure and takes the form
of a subtractive positive-semidefinite correction. We therefore decompose
\begin{equation}
\delta Q_{\mathrm{eff}}
=
-\zeta\,\Sigma(Q)
+
\delta Q_{\mathrm{res}},
\label{eq:deltaQeff}
\end{equation}
where $\Sigma(Q)$ is a positive semidefinite contribution of the form
\begin{equation}
\Sigma(Q) = B(Q) C(Q)^{-1} B(Q)^{\top} \geq 0,
\label{eq:SigmaDef}
\end{equation}
with $B(Q)$ and $C(Q)$ denoting the slow--fast and fast--fast blocks in the
decomposition \eqref{eq:quadraticBlock}. The coefficient $\zeta$ parametrizes
the effective strength of the elimination-induced coupling.

In this formulation, the Schur contribution $\Sigma(Q)$ is not a modeling
choice but a direct structural consequence of the underlying GKSL reduction.
Its dependence on $Q$ reflects the fact that the slow--fast decomposition,
and hence the induced coupling between sectors, is itself determined by the
instantaneous coarse-grained state. Accordingly, $\Sigma(Q)$ should be viewed
as a structural functional encoding the feedback of eliminated degrees of
freedom on the retained sector under iterated coarse graining.

The term $\delta Q_{\mathrm{res}}$ collects contributions not fixed by the
Schur structure. These include weak anisotropic driving, disorder in monitored
couplings, or deviations from idealized elimination. To leading order, its
trace component can be absorbed into an overall rescaling of $Q$, so that the
nontrivial contribution is taken to be traceless:
\begin{equation}
\delta Q_{\mathrm{res}}
=
a_k A_k,
\qquad
\mathrm{tr}(A_k)=0.
\label{eq:residualAniso}
\end{equation}
We emphasize that the detailed parametrization of this residual term is not
unique; the results below depend only on the presence of a generic traceless
perturbation competing with the Schur-induced contribution.

Iterating the coarse-graining step and introducing a normalization map
$\mathcal{C}$ that removes the overall scale yields the effective flow
\begin{equation}
Q_{k+1}
=
\mathcal{C}
\left(
Q_k
-
\zeta\,\Sigma_k
+
a_k A_k
\right),
\label{eq:flow}
\end{equation}
with $\Sigma_k=\Sigma(Q_k)\geq0$.

The update rule \eqref{eq:flow} should therefore be understood as the minimal
discrete-time dynamical system consistent with repeated GKSL-based
coarse-graining at the level of quadratic response tensors. The Schur term is
fixed by the elimination mechanism, while the anisotropic contribution
parametrizes residual degrees of freedom not determined by the reduction
alone.

\paragraph{Decomposition and signature sectors}

To characterize the resulting dynamics, we decompose the quadratic tensor into
isotropic and traceless components,
\begin{equation}
Q = q I + S, \qquad q = \frac{1}{d}\,\mathrm{tr}(Q), \quad \mathrm{tr}(S) = 0,
\label{eq:decomposition}
\end{equation}
where $d$ is the dimension of the slow sector. The scalar component $q$
controls the collective spectral shift, while $S$ encodes anisotropic
deviations.

Under the flow \eqref{eq:flow}, the positive semidefinite Schur contribution
$\Sigma_k$ shifts spectral weight downward in the directions coupled to
eliminated modes. Its isotropic part reduces the mean response, while its
anisotropic part depends on the structure of the slow--fast coupling.
The perturbations $A_k$ redistribute spectral weight without changing the trace
at leading order. The competition between these two effects organizes the
dynamics into signature sectors, classified by the number $n_-$ of negative
eigenvalues of $Q$.

Configurations with a single distinguished direction arise when the isotropic
component dominates over anisotropic fluctuations. A sufficient spectral
separation condition is
\begin{equation}
|q| > \|S\|_{\mathrm{op}},
\label{eq:lorentzCondition}
\end{equation}
where $\|\cdot\|_{\mathrm{op}}$ denotes the operator norm. In the convention
used below, the full tensor is represented as
$Q=\mathrm{diag}(q_N,Q_T)$ with $q_N>0$ fixed. When the tangential block has
three negative eigenvalues, the full tensor therefore has one positive and
three negative directions, corresponding to a Lorentz-type $(1,3)$ signature.
More generally, Eq.~\eqref{eq:lorentzCondition} expresses the stability of a
single distinguished sign sector against bounded anisotropic perturbations.

Conversely, when
\begin{equation}
|q| \lesssim \|S\|_{\mathrm{op}},
\label{eq:splitCondition}
\end{equation}
anisotropic fluctuations become comparable to the collective spectral shift
and multiple sign changes can occur, leading to split-signature
configurations.

\begin{figure}[t]
\centering
\includegraphics[width=\linewidth]{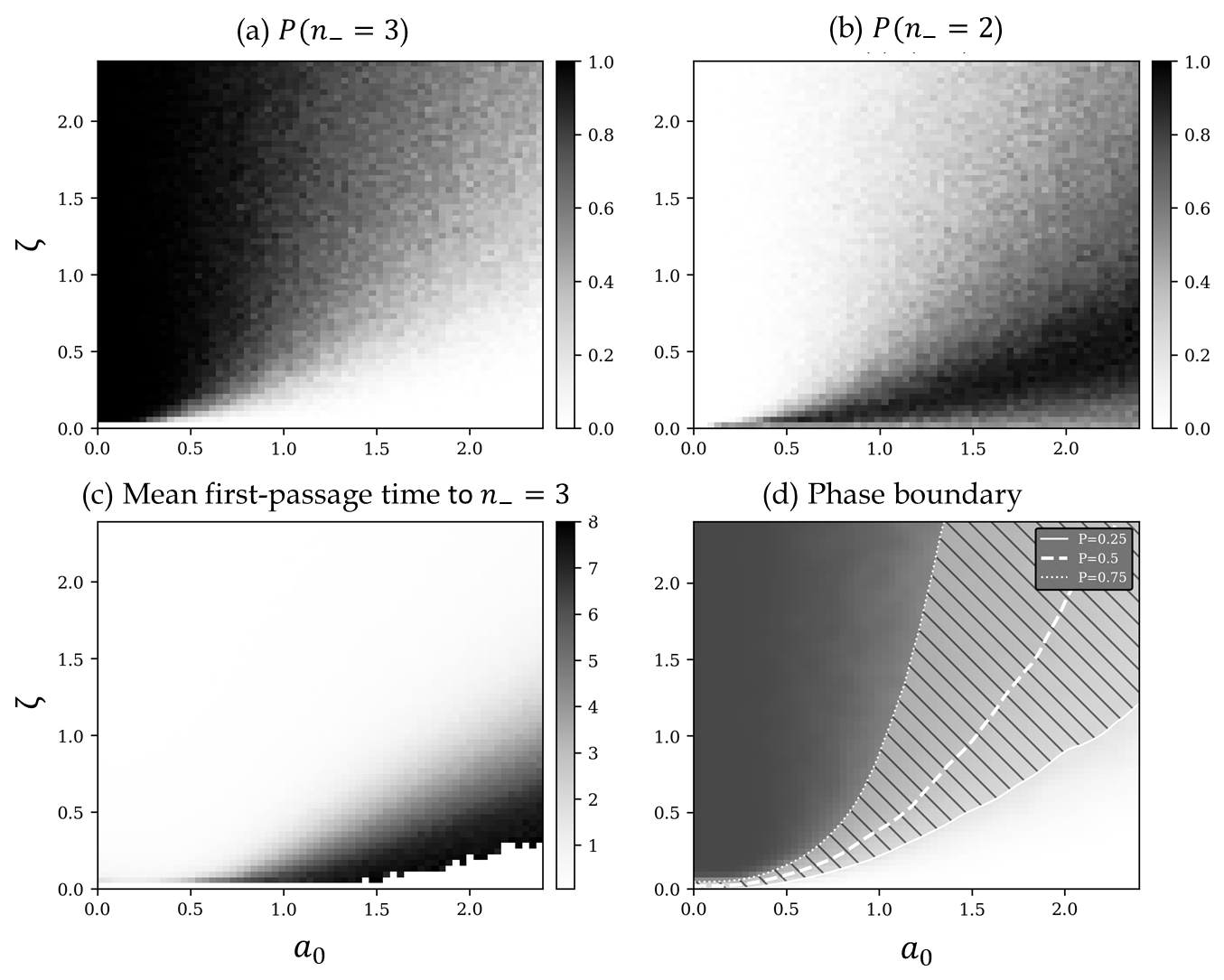}
\caption{
Statistical characterization of the signature dynamics under the coarse-grained flow.
(a) Probability $P(n_- = 3)$ of obtaining three negative eigenvalues.
(b) Probability $P(n_- = 2)$.
(c) Mean first-passage time to the $n_- = 3$ sector.
(d) Phase boundary extracted from probability contours.
The results show a well-defined crossover between regimes dominated by different signature sectors, controlled by the ratio of anisotropic driving to Schur-induced damping.
}
\label{fig:phase_structure}
\end{figure}

\paragraph{Numerical implementation.}
The statistical results shown in Fig.~\ref{fig:phase_structure} are obtained
from ensembles of discrete-time trajectories generated from the effective
update rule \eqref{eq:flow}. The simulations are performed in a reduced
representation in which the quadratic tensor is decomposed into a fixed
normal component and a dynamical tangential block,
\begin{equation}
Q = \mathrm{diag}(q_N, Q_T),
\end{equation}
with $q_N>0$ fixed and $Q_T \in \mathbb{R}^{3\times 3}$ evolved under the flow.

\begin{figure}[t]
\centering
\includegraphics[width=0.49\linewidth]{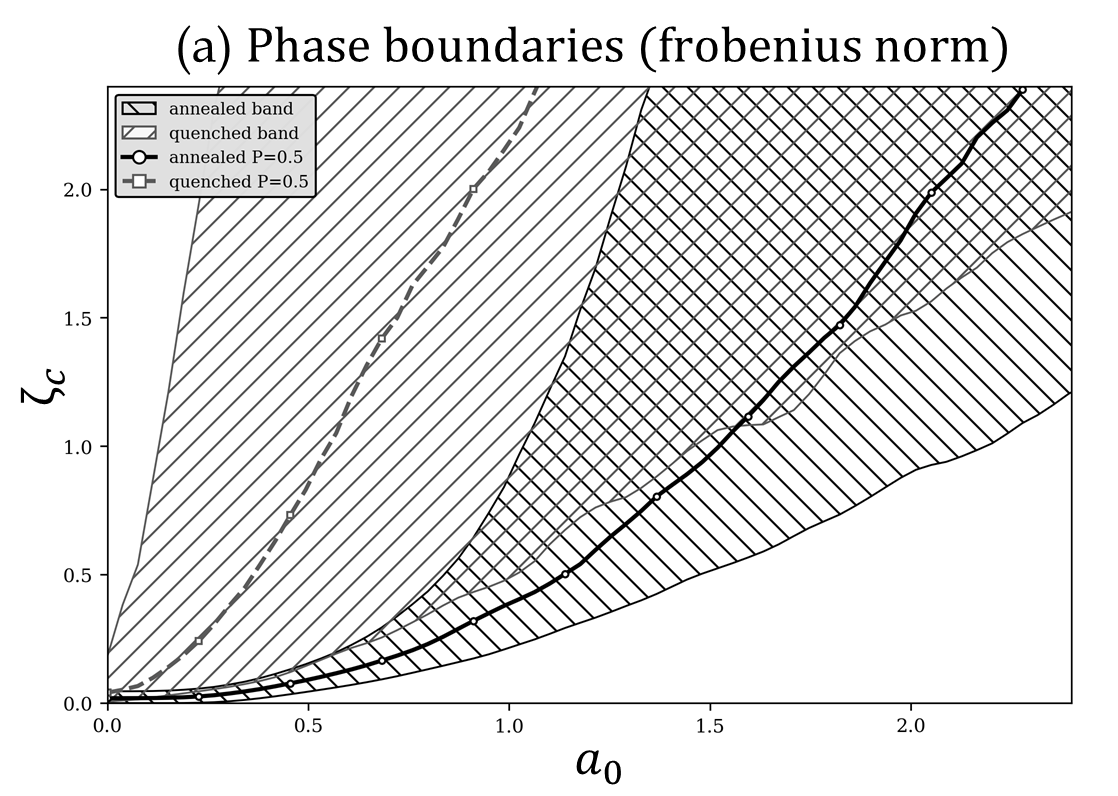}
\includegraphics[width=0.49\linewidth]{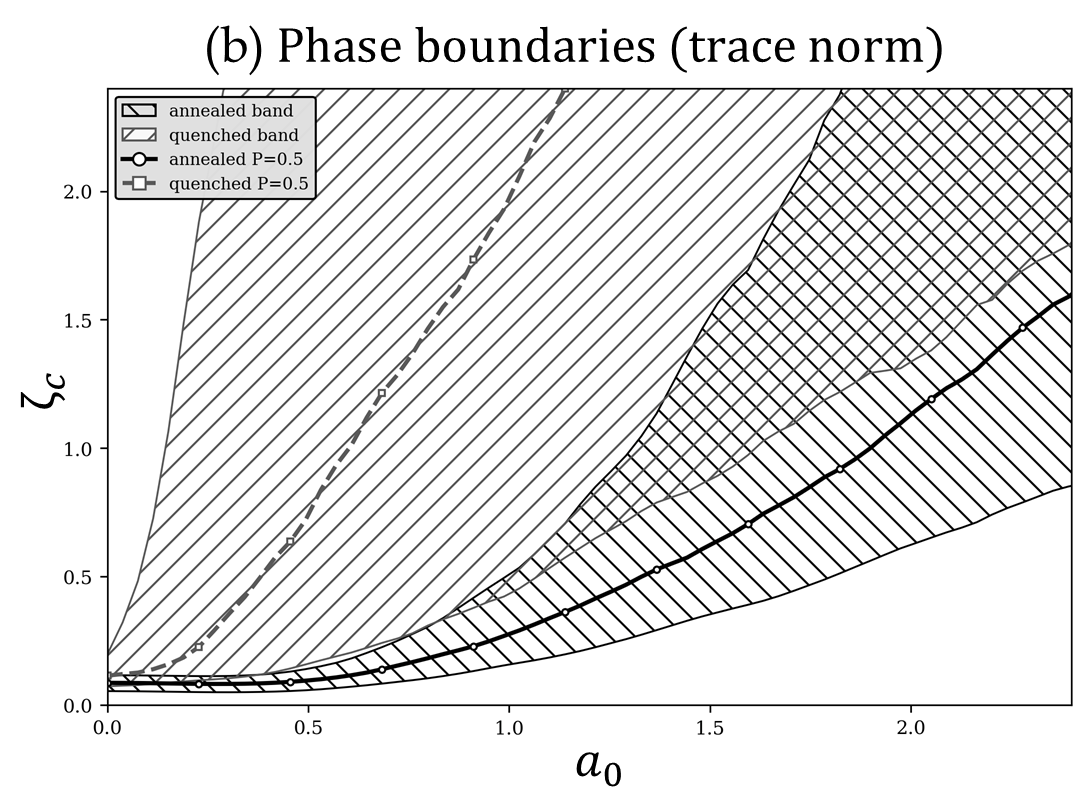}
\caption{Comparison of phase boundaries obtained under annealed and quenched disorder
for two different choices of norm used to quantify anisotropy.
(a) Frobenius norm (b) trace-based norm.
Shaded regions indicate parameter ranges in which the system transitions between
distinct signature sectors, and solid lines denote representative transition
curves (e.g.\ $P=0.5$).
While the precise location of the phase boundary is mildly shifted between the
two cases, its overall shape and organization remain unchanged, indicating that
the transition is robust within the effective model.
}
\label{fig:annealed_quenched}
\end{figure}

At each iteration step, the Schur contribution is implemented as a positive
semidefinite matrix of the form
\begin{equation}
\Sigma_k = B_k C_k^{-1} B_k^{\top},
\end{equation}
where the fast-sector response $C_k$ is sampled as a positive definite matrix
with a broad spectrum, and the coupling matrices $B_k$ are drawn from Gaussian
ensembles. In the data shown in the main text, two realizations of the
resulting positive-semidefinite structure are used: in
Fig.~\ref{fig:phase_structure} and Fig.~\ref{fig:annealed_quenched}(a), a
lognormal random model for $C_k$ is combined with Gaussian sampling of $B_k$,
while Fig.~\ref{fig:annealed_quenched}(b) uses an isotropic Wishart-type
construction,
\begin{equation}
\Sigma_k = G_k G_k^{\top},
\end{equation}
with $G_k$ a random Gaussian matrix. These implementations capture the
positive-semidefinite Schur structure while avoiding unnecessary
model-specific features.

The anisotropic perturbations $A_k$ are taken as random traceless symmetric
matrices with unit Frobenius norm, ensuring $\mathrm{tr}(A_k)=0$ and
$\|A_k\|_{\mathrm{F}}=1$. Their strength is controlled by a scalar parameter
$a_k = a_0 e^{-\beta \lambda_k}$, which decreases along the flow and models
the decay of anisotropic driving under coarse graining.

The effective update step is therefore implemented as
\begin{equation}
Q_T \;\mapsto\; \mathcal{C}\!\left(
Q_T - \zeta\, \Sigma_k + a_k A_k
\right),
\end{equation}
where $\zeta$ sets the relative strength of the Schur contribution. The
normalization map $\mathcal{C}$ rescales the tensor to a fixed reference
scale. In Fig.~\ref{fig:phase_structure}, a Frobenius normalization is used,
while in Fig.~\ref{fig:annealed_quenched} both Frobenius and trace-based
normalizations are considered for comparison. In all cases, the normalization
acts projectively and preserves the signature of the quadratic form.

Simulations are performed on parameter grids $(a_0,\zeta)$ using ensembles
of $\mathcal{O}(10^2)$ trajectories per point, each evolved over
$\mathcal{O}(10^2)$ coarse-graining steps. For each trajectory, we compute
the signature of the full tensor $Q$ and record the first-passage time to the
target sector $n_-=3$ of the tangential block. Since $q_N>0$ is fixed, this
sector corresponds to a Lorentz-type $(1,3)$ signature of the full tensor.
The reported observables correspond to ensemble averages of these quantities.

We have verified that the qualitative features shown in
Fig.~\ref{fig:phase_structure}, in particular the emergence of a connected
crossover region and the organization into signature sectors, are robust
under variations of the sampling scheme, including alternative positive-
semidefinite Schur models and different realizations of anisotropic
perturbations. These additional checks are not shown explicitly, as the
figures focus on the minimal implementation described above.

\paragraph{Statistical structure of the flow}

To probe the structure of the effective dynamics, we analyze ensembles of
trajectories generated by Eq.~\eqref{eq:flow}. For each realization, we
extract the signature of $Q_k$ and compute statistical observables, including
the probability $P(n_-)$ of obtaining a given number of negative eigenvalues.
This ensemble-based analysis provides a statistical characterization of the
effective coarse-grained dynamics, in analogy with nonequilibrium approaches
to irreversibility and transport \cite{Zwanzig2001,LandauLifshitzStat}.

The resulting statistical structure reveals a well-defined crossover boundary
separating regimes with distinct dominant signature behavior. As shown in
Fig.~\ref{fig:phase_structure}, the probability distribution concentrates in
well-defined sectors, indicating that signature is dynamically selected within
the effective model. The change between regimes is continuous and reflects a
reorganization of trajectories in parameter space rather than a singular
transition in the underlying microscopic dynamics.

We further consider first-passage observables associated with reaching a given
signature sector. Denoting by $\tau$ the first iteration at which a trajectory
enters a target sector, we define
\begin{equation}
\langle \tau \rangle = \mathbb{E}\!\left[\min\{k : Q_k \in \mathcal{S}_{\mathrm{target}}\}\right],
\label{eq:fpt}
\end{equation}
which exhibits a pronounced increase near the boundary between regimes. As
visible in Fig.~\ref{fig:phase_structure}(c), this behavior indicates a
slowing down associated with competing spectral configurations.

\paragraph{Annealed vs. quenched structure}

A finer characterization of the crossover structure is obtained by
distinguishing between annealed and quenched realizations of the anisotropic
driving. In the annealed case, the perturbations $A_k$ fluctuate independently
at each step, whereas in the quenched case they are fixed along a trajectory.

The qualitative structure of the crossover boundary is robust under variations
of the statistical properties of $A_k$, indicating that, within the class of
effective models considered, the observed organization reflects a structural
feature of the coarse-grained dynamics. As shown in
Fig.~\ref{fig:annealed_quenched}, temporal correlations in the driving
influence the sharpness and location of the crossover, but do not remove the
organization into signature sectors.

\paragraph{Summary}

Within the effective flow \eqref{eq:flow}, the Zeno--Schur mechanism induces a
nontrivial organization of quadratic response tensors, in which signature is
dynamically selected. The existence of negative directions in
$Q_{\mathrm{eff}}$ follows directly from the Schur-complement structure
derived in Sec.~\ref{sec:2}. By contrast, the detailed statistical
organization of signature sectors arises within the effective coarse-grained
model defined by Eq.~\eqref{eq:flow}. The relevant control parameter is the
ratio between anisotropic driving and Schur-induced damping. In the reduced
representation used here, the sector with $n_-=3$ tangential directions
corresponds to a Lorentz-type $(1,3)$ signature of the full quadratic tensor.

\section{Mechanism and infrared organization}
\label{sec:5}

To elucidate the origin of the signature structure identified in
Sec.~\ref{sec:4}, we analyze the effective dynamics in terms of the
decomposition \eqref{eq:decomposition} into isotropic and traceless
components. This representation makes explicit that the Zeno--Schur
mechanism induces two qualitatively distinct channels of evolution acting on
the quadratic response tensor.

The Schur-complement contribution $\Sigma(Q) \geq 0$ induces, to leading
order, a systematic reduction of the isotropic component $q$. In the absence
of anisotropic perturbations, repeated application of the update
\eqref{eq:flow} therefore shifts the spectrum of $Q$ collectively toward
smaller values. This behavior reflects the subtractive nature of the
elimination-induced renormalization derived in Sec.~\ref{sec:2}, and provides
a \emph{compressive channel} acting on the overall spectral scale.

By contrast, the traceless perturbations $A_k$ entering
Eq.~\eqref{eq:flow} do not affect $\mathrm{tr}(Q)$, but redistribute spectral
weight among eigenvalues. Their effect is to broaden the eigenvalue
distribution and enhance anisotropy without shifting the mean response. This
defines a complementary \emph{redistributive channel}, which competes with the
collective drift induced by the Schur term.

The dynamics of the effective model can therefore be understood as the
interplay between these two channels: a compressive isotropic flow generated
by Zeno--Schur elimination, and a redistributive anisotropic flow generated by
residual perturbations.

When the Schur-induced contribution dominates, the condition
\eqref{eq:lorentzCondition} is satisfied and the system approaches
configurations in which the spectrum is tightly clustered around a common
shift with bounded anisotropic deviations. In this regime, all but one
eigenvalue share a common sign, while a single direction is separated by the
collective drift. In the representation used in Sec.~\ref{sec:4}, this
corresponds to the stabilization of a Lorentz-type $(1,3)$ signature of the
full quadratic tensor. The concentration of trajectories in this sector is
visible in Fig.~\ref{fig:phase_structure}.

Conversely, when anisotropic contributions become comparable to or exceed the
Schur-induced drift, the condition \eqref{eq:splitCondition} applies and the
redistributive channel dominates. In this regime, spectral weight is
continuously transferred between eigenvalues, and multiple sign changes can
occur. The system then explores configurations with mixed or split
signatures, as observed in Sec.~\ref{sec:4}.

The crossover between these regimes occurs when the characteristic scales of
the two channels become comparable,
\begin{equation}
|q| \sim \|S\|_{\mathrm{op}}.
\end{equation}
Near this boundary, leading eigenvalues approach zero, resulting in enhanced
sensitivity to perturbations and a strong dependence on the realization of
anisotropic driving. This behavior is reflected in the increase of
first-passage times in Fig.~\ref{fig:phase_structure}(c), indicating a regime
of marginal stability rather than a sharp transition.

Importantly, this mechanism does not rely on fine-tuning of microscopic
parameters. It follows from the structural properties of Zeno-induced
elimination combined with generic anisotropic perturbations. The compressive
channel is fixed by the Schur structure and therefore universal within the
class defined in Sec.~\ref{sec:2}, while the redistributive channel captures
model-dependent but broadly generic anisotropic effects. The resulting
organization into signature sectors is therefore robust under variations of
the microscopic realization.

This robustness is further supported by the weak dependence on the choice of
anisotropy measure, demonstrated in Fig.~\ref{fig:annealed_quenched}. While the
precise location of the crossover boundary depends on details of the
perturbations, its qualitative structure remains unchanged.

Taken together, these results show that the signature of the quadratic
response tensor is not fixed a priori, but emerges dynamically from the
competition between elimination-induced compression and anisotropic
redistribution. Within the effective description, this mechanism provides a
natural explanation for the organization of trajectories into distinct
signature sectors observed in Sec.~\ref{sec:4}, including the stabilization of
Lorentz-type configurations in a finite parameter regime.

\section{Operational reconstruction and relation to effective dynamics}
\label{sec:6}

In this section, we clarify the operational meaning of the
Schur-renormalized quadratic tensor $Q_{\mathrm{eff}}$ and its relation to
experimentally accessible fluctuation observables. A key point is that
$Q_{\mathrm{eff}}$ arises as a response object in the effective coarse-grained
description, whereas experimentally accessible quantities are typically
defined through fluctuation statistics. Establishing the relation between
these objects therefore requires additional dynamical input.

As emphasized in Sec.~\ref{sec:2}, $Q_{\mathrm{eff}}$ is directly linked to
the reduced drift only when the slow dynamics admits a compatible local
quadratic response representation. Under this condition, $Q_{\mathrm{eff}}$
determines the effective drift structure on the slow manifold. In more
general nonequilibrium settings, it constrains the drift but does not
coincide directly with the curvature of the measured stationary distribution.

For clarity, we summarize the notation used in this section:
$Q_{\mathrm{eff}}$ denotes the Schur-renormalized quadratic response tensor,
$M$ the effective drift matrix of the reduced dynamics, and
$G_{\mathrm{eff}}$ the locally reconstructed curvature of the fluctuation
distribution.

\subsection{Effective drift from quadratic response}

Following Sec.~\ref{sec:2}, we assume that the Zeno-reduced slow dynamics
admits a local gradient or generalized Onsager representation. To leading
order in the slow variables, the effective response functional is
\begin{equation}
\Phi_{\mathrm{eff}}(p)
=
\frac{1}{2} p^{\top} Q_{\mathrm{eff}} p,
\label{eq:PhiEff}
\end{equation}
where $Q_{\mathrm{eff}}$ denotes the Schur-renormalized response tensor.

Under the additional assumption of a positive mobility matrix $\mu$ compatible
with the slow-sector variables, the deterministic component of the reduced
dynamics is
\begin{equation}
\dot p
=
- \mu \nabla_p \Phi_{\mathrm{eff}}(p)
=
- \mu Q_{\mathrm{eff}} p,
\label{eq:DriftFromQeff}
\end{equation}
where $\mu$ is inherited from the dissipative part of the reduced dynamics.

The linear drift matrix is therefore
\begin{equation}
M = \mu Q_{\mathrm{eff}}.
\label{eq:MfromQ}
\end{equation}

Equation~\eqref{eq:MfromQ} should be understood as a conditional
identification: it holds in regimes where the reduced dynamics admits the
gradient-response structure specified above. More generally,
$Q_{\mathrm{eff}}$ defines a quadratic response sector, while the drift must
be obtained from the reduced generator.

\subsection{Stochastic dynamics and stationary distribution}

Including fluctuations, the reduced dynamics may be represented locally by a
linear stochastic equation
\begin{equation}
\dot p
=
- M p + \xi,
\label{eq:Langevin}
\end{equation}
where $\xi$ is a noise term with covariance
\begin{equation}
\langle \xi_i(t)\xi_j(t') \rangle
=
2 D_{ij} \delta(t-t').
\label{eq:Noise}
\end{equation}

The associated probability density satisfies the Fokker--Planck equation
\begin{equation}
\partial_t P(p,t)
=
\nabla_p \cdot \left( M p\, P(p,t) + D \nabla_p P(p,t) \right).
\label{eq:FokkerPlanck}
\end{equation}

For a stable linear drift, the stationary covariance is determined by the
Lyapunov equation
\begin{equation}
M^{\top} \Gamma + \Gamma M = 2D,
\label{eq:Lyapunov}
\end{equation}
where $\Gamma = \langle p p^{\top} \rangle$ is the covariance matrix.

\subsection{Local response and logarithmic curvature}

Independently of global Gaussianity, the local fluctuation response can be
defined through the curvature of the log-probability,
\begin{equation}
G_{\mathrm{eff},ij}(p_0)
=
-
\left.
\frac{\partial^2}{\partial p_i \partial p_j}
\log P(p)
\right|_{p=p_0}.
\label{eq:Geff}
\end{equation}

The tensor $G_{\mathrm{eff}}$ characterizes the local quadratic structure of
the measured fluctuation distribution and is directly accessible from
experimental data. Importantly, $G_{\mathrm{eff}}$ is a local observable and,
in general nonequilibrium situations, does not coincide with either
$Q_{\mathrm{eff}}$ or the inverse covariance matrix $\Gamma^{-1}$.

\subsection{Relation between $Q_{\mathrm{eff}}$ and $G_{\mathrm{eff}}$}

Equations~\eqref{eq:MfromQ}--\eqref{eq:Lyapunov} show that, under the
gradient-response assumption, $Q_{\mathrm{eff}}$ enters the fluctuation
structure through the drift matrix $M$. The measured curvature
$G_{\mathrm{eff}}$, however, is determined jointly by the drift and the noise
matrix $D$.

Thus, in general nonequilibrium settings,
\begin{equation}
G_{\mathrm{eff}} \neq Q_{\mathrm{eff}},
\end{equation}
unless additional fluctuation--dissipation conditions are satisfied.
Nevertheless, a loss of positive definiteness of $Q_{\mathrm{eff}}$ modifies
the spectral structure of the drift matrix $M=\mu Q_{\mathrm{eff}}$
whenever the mobility $\mu$ is positive. This affects relaxation modes and
can induce measurable changes in fluctuation statistics, providing an
indirect operational signature of the Schur-induced instability.

\subsection{Special case: local Einstein relation}

A direct identification between $Q_{\mathrm{eff}}$ and the measurable
curvature arises if the reduced dynamics satisfies a local Einstein relation,
\begin{equation}
D = \beta^{-1} \mu,
\label{eq:Einstein}
\end{equation}
with $\beta$ an effective inverse temperature.

In this case, provided that the drift is of the gradient form
\eqref{eq:DriftFromQeff}, the stationary solution of
Eq.~\eqref{eq:FokkerPlanck} is
\begin{equation}
P_{\mathrm{st}}(p)
\propto
\exp\!\left[
-\frac{\beta}{2} p^{\top} Q_{\mathrm{eff}} p
\right],
\label{eq:Gaussian}
\end{equation}
and the logarithmic curvature reduces to
\begin{equation}
G_{\mathrm{eff}} = \beta Q_{\mathrm{eff}}.
\label{eq:GeffEqualsQ}
\end{equation}

Equation~\eqref{eq:GeffEqualsQ} provides a direct experimental route to
$Q_{\mathrm{eff}}$ in regimes where an effective fluctuation--dissipation
relation holds. In such cases, the signature of $Q_{\mathrm{eff}}$ is directly
reflected in the measurable fluctuation geometry.

In quantum kinetic frameworks such as the quantum linear Boltzmann equation,
the effective response tensor plays a closely related role, linking drift and
fluctuation structure at the coarse-grained level. In this sense,
$Q_{\mathrm{eff}}$ provides a natural extension of such response descriptions
to monitored and Zeno-reduced dynamics.

\subsection{Experimental reconstruction}

The above relations define a practical reconstruction protocol. After
preparing the system in the Zeno-reduced stationary regime, one measures the
coarse-grained variables $p_i$ and reconstructs the probability density
$P(p)$ or its local curvature via Eq.~\eqref{eq:Geff}. The resulting tensor
$G_{\mathrm{eff}}$ can then be compared to predictions based on
Eqs.~\eqref{eq:MfromQ}--\eqref{eq:Lyapunov}, provided the reduced drift and
noise matrices are independently characterized.

In systems where Eq.~\eqref{eq:Einstein} is approximately satisfied,
$Q_{\mathrm{eff}}$ can be obtained directly from curvature measurements.
Outside this regime, one reconstructs $M$ and $D$ separately and infers the
response tensor only after validating the gradient-response structure.

\subsection{Connection to minimal realizations}

The synthetic-frequency implementation discussed in Sec.~\ref{sec:3} provides
a natural platform for this protocol. In such systems, the relevant
quadratures can be monitored, and the fluctuation statistics reconstructed
from repeated measurements. As shown in that example, tuning the
coherence-sensitive coupling controls the magnitude of the Schur correction
and drives the system across the threshold where the effective quadratic
structure changes its signature.

\medskip

Taken together, these results establish a conditional but operationally
testable connection between the Schur-renormalized tensor $Q_{\mathrm{eff}}$
and experimentally accessible fluctuation observables. The connection is
direct in systems satisfying a local Einstein relation and indirect otherwise,
through the measured drift and noise matrices.

Finally, the ensemble simulations discussed in Sec.~\ref{sec:4} identify
regions in parameter space in which the effective flow enters a Lorentzian
signature sector\footnote{We emphasize that the terminology ``Lorentzian''
refers here solely to the signature $(1,d-1)$ of the quadratic form and does
not imply any a priori geometric or spacetime interpretation.} with a finite
spectral margin. Stability of this sector follows from standard eigenvalue
perturbation theory: if the tangential block satisfies
\begin{equation}
\lambda_{\max}(Q_{\mathrm{tan}}) \leq -\Delta < 0,
\label{eq:SpectralGap}
\end{equation}
then any anisotropic perturbation $A$ with
\begin{equation}
\|A\|_{\mathrm{op}} < \Delta
\label{eq:AnisotropyBound}
\end{equation}
preserves the number of negative tangential eigenvalues. Thus the Lorentzian
sector is locally stable whenever it is reached with a finite gap to the
signature-changing boundary.

These considerations show that the signature structure of quadratic response
tensors is, in principle, experimentally accessible in monitored systems,
provided the relation between response, drift, and fluctuation curvature is
properly characterized.

\section{Conclusion}
\label{sec:7}

We have analyzed the structure and dynamics of quadratic response tensors in
monitored open quantum systems subject to strong measurement backaction.
Within the GKSL framework and under standard assumptions of time-scale
separation and quadratic closure, we have shown that Zeno-induced elimination
of fast degrees of freedom generates a subtractive renormalization with
Schur-complement structure.

This mechanism has a direct structural consequence: positive definiteness of
quadratic response tensors is not generally preserved under coarse graining.
Instead, coupling between slow and rapidly damped sectors induces a
systematic reduction of the effective response, which can generate negative
directions even when the microscopic tensor is strictly positive definite.
This identifies a general pathway by which indefinite quadratic structure can
emerge in Zeno-reduced open quantum dynamics.

To investigate the implications of this mechanism, we introduced a minimal
effective coarse-grained flow on the space of quadratic tensors, in which the
Schur-induced compressive contribution competes with anisotropic
perturbations. Within this framework, the dynamics organizes into distinct
signature sectors characterized by the number of negative eigenvalues. The
resulting behavior exhibits a robust crossover between regimes dominated by
different signatures, accompanied by enhanced first-passage times near the
crossover boundary.

At the mechanistic level, this behavior can be understood as the interplay
between two generic channels: a compressive isotropic flow induced by the
Schur term, and a redistributive anisotropic flow arising from residual
perturbations. Their competition provides a simple and robust explanation for
the observed organization of signature sectors, without requiring fine-tuning
of microscopic parameters.

Our results show that the signature of effective quadratic response tensors
in monitored open systems is not fixed a priori, but can be dynamically
modified by measurement-induced coarse graining. In this sense, the
Zeno--Schur mechanism defines a structural constraint on the space of
admissible effective response tensors, within which dynamical processes select
specific signature sectors.

The operational analysis in Sec.~\ref{sec:6} further shows that this structure
is, in principle, experimentally accessible. In regimes satisfying an
effective fluctuation--dissipation relation, the Schur-renormalized tensor can
be directly reconstructed from fluctuation measurements. More generally, its
signature influences the drift structure and can be inferred indirectly from
measured relaxation and fluctuation properties.

Within the effective description considered here, a Lorentz-type signature
sector emerges as a stable configuration in a finite parameter regime. We
emphasize that this terminology refers solely to the signature structure of
the quadratic form and does not imply any a priori geometric or spacetime
interpretation. Rather, it characterizes a dynamically selected configuration
of the effective response tensor.

Compared to standard treatments of adiabatic elimination in open quantum
systems, which focus primarily on effective generators, the present work
highlights the induced transformation of quadratic response structures. In
contrast to Gaussian open-system analyses, where positivity of covariance-like
objects is typically preserved, the Schur-complement mechanism provides a
systematic route to indefinite quadratic forms. This identifies a qualitatively
distinct aspect of monitoring-induced coarse graining that is not captured by
conventional equilibrium or linear-response frameworks.

More broadly, the emergence and stabilization of nontrivial signature sectors
suggest that effective kinematic structures can arise dynamically from
monitoring-induced constraints. While possible connections to approaches in
which kinematic structure emerges from more fundamental degrees of freedom
\cite{Ambjorn2001CDT,Ambjorn2004CDT,Sorkin2003CausalSet} are intriguing, such
interpretations remain speculative within the present framework.

Future directions include deriving the effective flow from explicit
microscopic models, extending the analysis beyond quadratic closure, and
exploring experimental platforms in which the predicted signature
reorganization can be probed. In particular, connections to quantum kinetic
descriptions, such as the quantum linear Boltzmann equation
\cite{Vacchini2009}, and to recent developments in Zeno-constrained dynamics
\cite{Pernice2026} provide promising routes toward a more complete microscopic
understanding.

\begin{appendices}

\appendix

\end{appendices}

\newpage

\bibliography{signature_structure}

\end{document}